\documentclass[twocolumn,showpacs,prb,amsmath,amssymb]{revtex4}

\usepackage{graphicx}
\usepackage{bm}
\usepackage{mathcomp} 

\begin{document}

\title{
Magnetic and transport percolation 
 in diluted magnetic semiconductors
}

\author{A. Kaminski}
\author{S. \surname{Das Sarma}}
\affiliation{
Condensed Matter Theory Center,
Department of Physics, University of Maryland, College Park, Maryland
20742-4111}

\begin{abstract}
The ferromagnetic transition in a diluted magnetic semiconductor with
localized charge carriers is inevitably a percolation transition. In
this work we theoretically study the correlation between this magnetic
percolation and transport properties of the sample, including the
possibility of a simultaneous transport percolation.  We find
nontrivial signatures of the percolating magnetic clusters in the
transport properties of the system, including interesting
non-monotonic temperature dependence of the system resistivity.
\end{abstract} 


\maketitle

\section{Introduction}

``Diluted magnetic semiconductors'' (DMS) is a common name for a wide
class of semiconductor materials doped with magnetic impurities (with
their relative concentration up to a few percent) which become
ferromagnetic at low temperatures. Magnetic semiconductors are now
actively investigated both theoretically and experimentally, due to
their potential applications in new generations of semiconductor
devices.

An example of a DMS material is given by Ga$_{1-x}$Mn$_x$As (GaAs
doped with Mn). For the ferromagnetic transition to occur, the
concentration of Mn impurities should be relatively high, $x\sim 0.01
- 0.05$. Such high concentration of impurities makes growth of
magnetic Ga$_{1-x}$Mn$_x$As a difficult problem, which was solved only
relatively recently.\cite{Ohno96} In order to prevent phase
separation, magnetic Ga$_{1-x}$Mn$_x$As should be grown at low
temperatures ($T\sim 200 - 300 \tccentigrade$), which results in
abundance of different types of crystal defects. These problems are
not specific to Ga$_{1-x}$Mn$_x$As, in fact, all presently known
magnetic semiconductors suffer from the same problems. As a result,
theoretical study of magnetic semiconductors is very difficult: two
factors -- strong disorder and exchange interaction (which causes
ferromagnetic transition) -- must be taken into account
non-perturbatively.

The mechanism of ferromagnetic transition is common for most (if not
for all) magnetic semiconductor systems. Since the concentration of
magnetic impurities is only of the order of few percent, their direct
exchange interaction does not play the decisive role in determining
the ground state of the system. The ferromagnetic transition occurs
due to interaction between impurity spins mediated by charge carriers,
which are present in appreciable quantities because of the high
concentration of defects in the crystal lattice. This picture seems to
be accepted more or less universally, but the detailed nature of the
transition varies greatly from a system to a system. Because of the
large number of different magnetic semiconductor materials, high
defect concentration in them, and possible variations of growth
processes, it is highly unlikely that a ``universal theory of magnetic
semiconductors'' will ever be developed. A theory limited to a certain
subclass of magnetic semiconductor materials, which share same basic
properties, is more likely to provide useful information on the nature
of the processes underlying the ferromagnetic transition and give some
insight into experimental results.

In this paper, we study ferromagnetic semiconductors that in the limit
of zero temperature become insulating. The mechanism of ferromagnetic
transition for an extreme case of such a system, when the charge
carriers are pinned to some point defects, was considered in our
earlier paper.\cite{KaminskiDasSarma2002} Here we include transport
properties in our consideration, and also extend our scope to the
systems where the charge carries are localized within extended regions
rather than pinned with the wave function decaying exponentially from
the localization center.

A characteristic feature of DMS physics is the presence of a
metal-insulator transition as a function of the magnetic impurity
concentration $x$. Typically for low $x$ ($\lesssim 0.03$) the system
(e.g. Ga$_{1-x}$Mn$_x$As) is an insulator with strongly localized
carriers and with the resistivity increasing exponentially as $T\to
0$. For larger $x$ the system tends to be ``weakly'' metallic in the
sense that the resistivity, although very large ($\sim 0.01 \ 
\Omega\cdot\textrm{cm}$), does not exhibit strong temperature
dependence. There are reports\cite{Matsukura98} of a reentrant
insulating phase at higher Mn concentration, $x>0.05$ (More recent
experiments \cite{Potashnik02} on carefully annealed samples do not
find this reentrant insulating behavior at higher Mn concentration.)
In all situations both metallic and insulating phases are found to be
ferromagnetic.  Other DMS systems show typically the same behavior
(i.e. the existence of a metal-insulator transition with a weakly
metallic state, which has a very high resistivity, for higher values
of $x$) or often just an insulating ferromagnetic state without any
metallic behavior whatsoever. Systems showing a metal-insulator
transition tend to have higher ferromagnetic transition temperature
$T_c$ in their metallic phases, and therefore understanding the
insulating ferromagnetic DMS phase takes on significance, not just
from the perspective of fundamental physics involving interplay of
disorder and magnetism, but also from the practical technological
motivation of enhancing $T_c$ in DMS materials. This is particularly
true since the current zeroth-order theoretical understanding of the
optimally doped metallic DMS ferromagnetism is based on a simple
continuum virtual crystal approximation Weiss mean-field theory
treatment of an RKKY-Zener model\cite{KonigReview01} that is
intrinsically invalid in the insulating regime. In fact, it is
currently unclear whether the observed DMS metal-insulator transition
is a disorder-driven impurity band or valence band transition. It is,
however, clear that disorder plays a critical role in DMS physics,
both in its transport properties and in the magnetic properties. We
focus in this work on understanding DMS transport properties in the
localized insulating regime by applying our recently developed
theory,\cite{KaminskiDasSarma2002} which has been successful in
explaining the magnetic properties on the insulating side of the
metal-insulator transition. We hope to elucidate the role of disorder
and magnetic clustering on DMS transport properties through this
theoretical study. Although the DMS systems with the highest $T_c$ do
not belong to this class, their resistivity is still very large, so
disorder must play a very important role in the properties of these
systems. Study of DMS systems that are insulating at $T\to0$, as done
in this paper, can shed some light on this role, which is difficult to
understand using the approach based on the picture of free charge
carriers.

So the systems we consider in this paper are insulating at $T=0$,
which means that the Fermi level is below the mobility edge in the
impurity band or below the bottom of the conduction band (for
electrons; for holes it would be the valence band) if there are no
extended states in the impurity band. From now on, we will refer to
this class of magnetic semiconductor systems as ``systems with
localized charge carriers.''  Taking into account the high defect
concentration in a typical magnetic semiconductor material, we must
conclude that the charge carrier density in systems with localized
charge carriers is highly inhomogeneous. Since it is charge carriers
that transmit exchange interaction between magnetic impurities, this
effective interaction must also be highly inhomogeneous. Thus, when
the temperature is lowered, the ferromagnetic transition will first
occur locally, in the regions with higher charge-carrier density, i.e.
with stronger effective exchange interaction between magnetic
impurities. As the temperature goes further down, these finite-size
regions, which have random sizes and positions, will grow and merge,
until finally the ferromagnetic correlation is established across the
whole sample.\cite{griffiths} Such a scenario, which must hold for any
magnetic semiconductor with localized carriers, implies that
ferromagnetic transition is inevitably a percolation transition, with
clusters of ferromagnetic regions already existing at $T>T_c$ and an
infinite magnetic cluster opening up at $T=T_c$. This percolation
picture of ferromagnetism in insulating DMS systems is now
well-accepted and has been verified\cite{Dagotto02,YangMacDonald02} via
direct numerical simulations.

The foundation for the study of ferromagnetic percolation transition
was laid down in the 70s for diluted ferromagnetic alloys, where
magnetic atoms interact directly with each
other.\cite{KorenblitEtAl73} Early numerical simulations of localized
DMS systems \cite{WanBhatt00} already indicated indirect evidence for
magnetic percolation playing a role in the magnetic transition.  Our
earlier paper\cite{KaminskiDasSarma2002} extended results of
Ref.~\onlinecite{KorenblitEtAl73} to magnetic semiconductor systems,
where the exchange interaction between magnetic atoms is not direct,
but is rather mediated by charge carriers.

The word ``percolation'' by itself assumes some transport. Even though
the mathematical percolation theory with its physical applications
goes beyond this literal meaning (an example is given by diluted
ferromagnetic alloys of Ref.~\onlinecite{KorenblitEtAl73}, where no
transport is considered), one may still naturally wonder if a
percolation transition in a given system means some enhancement of
charge transport. In ferromagnetic semiconductors, this question is
quite relevant, since charge carriers, which have some ability to move
around the sample, are present, and since the ferromagnetic
percolation transition is facilitated by these very carriers. In fact,
some experiments on magnetic semiconductors did observe connections
between ferromagnetic transition and enhancement of charge transport,
up to the point of the temperature dependence of resistivity being
non-monotonic with the maximum around the ferromagnetic transition
temperature.\cite{Matsukura98,Potashnik02,Tanaka2003} In
addition, the existence of the metal-insulator transition in DMS
systems makes transport percolation considerations relevant since
there is an intrinsic connection between metal-insulator transition
and percolation. We emphasize in this context that the magnetic
percolation transition, considered in
Ref.~\onlinecite{KaminskiDasSarma2002}, is purely a statement on the
percolation properties of the temperature-dependent magnetic clusters
in DMS materials, and has little to do (in a direct sense) with
transport percolation properties (which must take into account the
site-to-site hopping of the localized charge carriers, as we do in
this paper) -- it is, in principle, entirely possible for a system to
have a magnetic percolation transition with no transport percolation
altogether. In fact, these are two separate physical phenomena whose
relationship is theoretically explored in this paper in the specific
context of DMS systems.

The main purpose of this work is to establish connection between
ferromagnetic transition in a magnetic semiconductor system with
localized carriers (which, as we mentioned above, must have percolation
nature) and transport properties of such a system. The outline of the
paper is as follows: In Sec.~\ref{sec:model}, we introduce our model.
In Sec.~\ref{sec:strongly}, we consider the case of charge carriers
strongly pinned to localization centers. The ferromagnetic properties
in such a system were studied in our earlier
papers;\cite{KaminskiDasSarma2002,DasSarmaHwangKaminski2002} here we
discuss its transport properties. The considerations of this section
are based on the impurity positions being uncorrelated. The case of
impurities forming clusters in considered in Sec.~\ref{sec:clusters}.
Finally, in Sec.~\ref{sec:verge} we qualitatively discuss the
transport properties of a system, whose charge carriers are on the
verge of being delocalized. For all these systems, we also discuss
dependence of the resistivity on the applied magnetic field.
 
\section{Model Hamiltonian}
\label{sec:model}

Charge carriers in Ga$_{1-x}$Mn$_x$As are holes, donated by Mn
impurities. While other magnetic semiconductors may have electrons as
the carriers mediating exchange interaction between magnetic
impurities, in this paper we will refer to the charge carries as
``holes,'' just for the sake of brevity. All our conclusions would
hold no matter what the charge carriers are.

In fact, the whole model system we use is based on Ga$_{1-x}$Mn$_x$As,
where Mn atoms act both as magnetic impurities and acceptors, thus
providing both local moments, which order ferromagnetically at
$T<T_c$, and charge carriers, which mediate ferromagnetic interaction
between these local moments. We do not expect our results to be
specific to this particular model as far as the qualitative results
are concerned, so our conclusions should be valid for other systems
carrier-mediated insulating DMS as well.

In our model, a Mn impurity in Ga$_{1-x}$Mn$_x$As is presented by two
spin-degenerate levels, which we will refer to as ``the deep level''
and ``the shallow level.'' The deep level, when occupied by one
electron, provides the impurity's spin. Coulomb repulsion prevents a
second electron from entering this level. The shallow level plays the
role of the impurity's acceptor level. Since the electron wave
function of this level has larger localization radius, the Coulomb
interaction between a deep-level electron and a shallow-level one is
weaker than that between two electrons on the deep level.

The full Hamiltonian of the system reads:
\begin{eqnarray}
\label{H0}
\hat{H}_0&=&
\sum_{m}\left[\vphantom{\sum_{\alpha\beta}}
\sum_\alpha (-\varepsilon_d) a^\dagger_{m\alpha}
a^{\vphantom{\dagger}}_{m\alpha} 
+ U_d a^\dagger_{m\uparrow} a^{\vphantom{\dagger}}_{m\uparrow}
a^\dagger_{m\downarrow} a^{\vphantom{\dagger}}_{m\downarrow}
\right. \nonumber\\
&&\qquad+
\sum_\alpha (-\varepsilon_0) c^\dagger_{m\alpha}
c^{\vphantom{\dagger}}_{m\alpha} 
+ U_0 c^\dagger_{m\uparrow} c^{\vphantom{\dagger}}_{m\uparrow}
c^\dagger_{m\downarrow} c^{\vphantom{\dagger}}_{m\downarrow}
 \nonumber\\
&&\left.\qquad+
\sum_{\alpha\beta} U_{0d}
c^\dagger_{m\alpha} c^{\vphantom{\dagger}}_{m\alpha}
a^\dagger_{m\beta} a^{\vphantom{\dagger}}_{m\beta}\right]
 \nonumber\\
&+&
\sum_{mn\alpha}t^{\vphantom{\dagger}}_{mn} c^\dagger_{m\alpha}c^{\vphantom{\dagger}}_{n\alpha} 
\nonumber\\
&+&\sum_{mn\alpha}\left(t^{(0d)\vphantom{\dagger}}_{mn} 
c^\dagger_{m\alpha}a^{\vphantom{\dagger}}_{n\alpha} 
+ \textrm{h.c.}\right)\;,
\end{eqnarray}
where indices $m$ and $n$ run over magnetic impurities,
$a^\dagger_{m\alpha}$ and $c^\dagger_{m\alpha}$ are the creation
operators for an electron with spin $\alpha$ localized at the $m$th
impurity at the deep/shallow level respectively. The first and the
third terms in the brackets in Eq.~(\ref{H0}) represent the energies
of holes occupying the deep and the shallow level respectively. The
second and the fourth terms in the brackets describe the Coulomb
repulsion between two holes on each of these two levels, while the
fifth term is for the inter-level Coulomb repulsion. The first term
after the brackets accounts for hole hopping between the shallow levels
of two impurities.

The last term of Eq.~(\ref{H0}) describes hopping from the deep
level of one impurity to the shallow level of another.  Hopping
between deep levels is neglected because of the rapid fall-off of the
electron wave functions at these levels.  We take the Fermi energy
equal to zero, $\varepsilon_F\equiv 0$ throughout the paper.

The parameters of Hamiltonian (\ref{H0}) must be chosen to mimic
magnetic impurities in Ga$_{1-x}$Mn$_x$As. The deep level in the
ground state must be taken by only one hole, so we take
$U_d>\varepsilon_d>0$. The shallow, acceptor levels of the impurities
are to donate, on average, less than one hole per impurity, so the
Fermi level coincides with the energy of a hole placed onto the
shallow level of an impurity, whose lower level level is already taken
by one hole, which means that $U_{0d}-\varepsilon_0 =
\varepsilon_F\equiv 0$.  The bottom of the conduction band is assumed
to be separated from the impurity level $U_{0d}-\varepsilon_0$ by the
energy $E_0 \gg T, t_{mn}$, so the (delocalized) valence band states
can be safely excluded from our consideration.

We assume the wave functions of these localized levels to fall off
exponentially, with the characteristic decay radii equal to $a_d$ and
$a_0$ for deep and shallow levels respectively, with $a_d \ll a_0$.
Then the hopping matrix elements
$t^{\textrm{(0d)}\vphantom{\dagger}}_{mn}$ and
$t^{\vphantom{\dagger}}_{mn}$, which are proportional to the overlap
of the corresponding wave functions, are given by
\begin{subequations}
\begin{eqnarray}
t^{\vphantom{\dagger}}_{mn} &=&
t^{\vphantom{\dagger}}_0 \left(1+\frac{r_{mn}}{a_0}\right)
\exp\left(-\frac{r_{mn}}{a_0}\right)\ ,
\label{t0decay}\\
t^{(0d)\vphantom{\dagger}}_{mn}& =&
t^{(0d)\vphantom{\dagger}}_0 
\exp\left(-\frac{r_{mn}}{a_0}\right)\ .
\label{tddecay}
\end{eqnarray}
\end{subequations}

With these parameters, we may make the Schrieffer-Wolff transformation
\cite{SchriefferWolff66} for the deep levels, thus reducing
Hamiltonian (\ref{H0}) to
\begin{eqnarray}
\label{H1}
\hat{H}&=&
\sum_{m}\left[\sum_\alpha (-\varepsilon_0) c^\dagger_{m\alpha}
c^{\vphantom{\dagger}}_{m\alpha} 
+ U_0 c^\dagger_{m\uparrow} c^{\vphantom{\dagger}}_{m\uparrow}
c^\dagger_{m\downarrow} c^{\vphantom{\dagger}}_{m\downarrow}
\right] \nonumber\\
&+&
\sum_{mn\alpha}t^{\vphantom{\dagger}}_{mn} c^\dagger_{m\alpha}c^{\vphantom{\dagger}}_{n\alpha} 
\nonumber\\
&+&\sum_{lmn\alpha\beta}
J^{l\vphantom{\dagger}}_{mn}
\left(\textbf{S}_l\cdot \bm{\sigma}_{\alpha\beta}\right) 
c^\dagger_{m\alpha}c^{\vphantom{\dagger}}_{n\beta} 
\;,
\end{eqnarray}
where $\bm{\sigma} \equiv (\sigma_x, \sigma_y, \sigma_z)$ is the
vector of Pauli matrices,
\begin{eqnarray}
J^{l\vphantom{\dagger}}_{mn}& = &
t^{(0d)\vphantom{\dagger}}_{ml}t^{(0d)\vphantom{\dagger}}_{ln}
\left(\frac{1}{\varepsilon_d}+\frac{1}{U_d-\varepsilon_d}\right)
\nonumber\\
&=& \left(t^{(0d)\vphantom{\dagger}}_0 \right)^2
\left(\frac{1}{\varepsilon_d}+\frac{1}{U_d-\varepsilon_d}\right)
\exp\left(-\frac{r_{ml}+r_{ln}}{a_0}\right)
\;,
\label{Jlmn}
\end{eqnarray}
and 
\begin{equation}
\textbf{S}_l \equiv \sum_{\alpha\beta}
 \bm{\sigma}_{\alpha\beta}
a^\dagger_{l\alpha}a^{\vphantom{\dagger}}_{l\beta} 
\label{Sdefinition}
\end{equation}
The other parameters of the system are the concentration of impurities
$n_{\textrm{i}}$ and the concentration of holes $n_{\textrm{h}}$. The
spatial distribution of impurities is assumed to be random throughout
the paper, although not necessarily uncorrelated. We also assume, as
it is the case in Ga$_{1-x}$Mn$_x$As and many other DMS, that the
system is very heavily compensated, $n_{\textrm{h}}\ll
n_{\textrm{i}}$, with many more magnetic impurities that charge
carriers present in the system.

Depending on the values of the system parameters, the system described
by Hamiltonian (\ref{H1}) may be either insulating or metallic in the
limit $T\to 0$. In this paper we will concentrate on the former case.
Ferromagnetic transition in such a system inevitably has percolation
nature, as we pointed out in the introduction.

\section{Strongly localized charge carriers}
\label{sec:strongly}

In this section, we consider the case of $a_0^3 n_{\textrm{i}} \ll 1$,
with the impurity positions being uncorrelated.  Randomness in the
impurity positions leads to randomness in the hopping matrix elements
$t_{mn}$, which exponentially depend on the distances $r_{mn}$ between
the corresponding impurities. This strong (by orders of magnitude,
provided $a_0^3 n_{\textrm{i}} \ll 1$) variation of $t_{mn}$ creates
strong variation of localization potentials for the holes, which thus
will be tied to the pairs of least-separated
impurities.\cite{EfrosShklovskiiBook} We must note that the concrete
mechanism of hole localization does not have crucial impact on the
derivations and conclusions presented in this section.  In order to
keep our presentation straightforward we limit ourselves to the
framework of Hamiltonian~(\ref{H1}), when the holes are localized at
these closest impurity pairs. Any additional disorder can be taken into
account by inclusion of new terms to Eq.~(\ref{H1}) and proper
modification of the hole localization parameters.

With strongly localized holes, for study of magnetic properties of the
system Hamiltonian (\ref{H1}) can be reduced to:
\begin{equation}
\label{Hloc}
\hat{H}= \sum_{mj}J_{mj}\hat{\mathbf{S}}_m
\hat{\mathbf{s}}_j,
\end{equation}
where indices $m$ and $j$ label magnetic impurities and holes
respectively, and $\hat{\mathbf{S}}_m$/$\hat{\mathbf{s}}_j$ are the
impurity/hole spin operators. The expression for the matrix
elements $J_{mj}$ of the impurity-hole exchange interaction depend on
the hole localization mechanism. In the case of localization by 
the pairs of least-separated impurities,  it is given by
\begin{equation}
\label{Jpair}
J_{mj}=\frac12\left(J_{mn_j^{(1)}}+J_{mn_j^{(2)}}\right)\;
\end{equation}
for a hole localized at the impurities labelled by indices
$n_j^{(1)}$ and $n_j^{(2)}$.

Since the concentration of holes is much smaller than that of
impurities, $n_{\textrm{h}} \ll n_{\textrm{i}}$, one localized hole is
surrounded by many magnetic impurities.  Exchange interaction between
the (localized) holes and magnetic impurities leads to their mutual
polarization when temperature $T$ is below exchange
constant $J_{mj}$. Since exchange coupling $J_{mj}$
decays with the impurity-hole distance $r_{mj}$, see
Eq. (\ref{tddecay}) and (\ref{Jlmn}), the first impurities to get
their spins aligned with a hole's spin as the temperature decreases
are the ones most close to the hole's localization site. At lower
temperatures, more distant impurities have their spins polarized by
the hole to whose domain they belong. This complex consisting of a
hole and magnetic impurities polarized by it is conventionally called
``bound magnetic polaron.'' The characteristic radius size of a
polaron grows as the temperature decreases:\cite{KaminskiDasSarma2002}
\begin{equation}
\label{rcorr}
r_{\textrm{pol}}(T) = \frac{a_0}{2} 
\ln\frac{AJ_0\sqrt{a_0^3 n_{\textrm{i}}}}{T}\,,
\end{equation}
where $A\sim 1$ and
\begin{displaymath}
J_0 \equiv \left(t^{(0d)\vphantom{\dagger}}_0 \right)^2
\left(\frac{1}{\varepsilon_d}+\frac{1}{U_d-\varepsilon_d}\right)\;,
\end{displaymath}
cf.~Eq.~(\ref{Jlmn}).
As polarons overlap, they form polaron clusters with all
impurities belonging to a given cluster having their spins aligned in
the same direction.  The temperature\cite{KaminskiDasSarma2002} 
\begin{equation}
\label{Tc}
T_c\sim sS|J_0|  \left(a_0^3n_{\textrm{h}}\right)^\frac{1}{3}\! 
\sqrt{n_{\textrm{i}}/n_{\textrm{h}}}\
\exp\left(-\frac{0.86}{\left(a_0^3n_{\textrm{h}}\right)^\frac{1}{3}}\right)
\end{equation}
at which the infinite cluster
spanning the whole sample appears is the ferromagnetic transition
temperature.  The quantitative theory of this transition was recently
developed and presented by us in
Ref.~\onlinecite{KaminskiDasSarma2002} and further developed in
Ref.~\onlinecite{DasSarmaHwangKaminski2002}. In this paper we
concentrate on transport properties of such a system and their
correlation with the magnetic properties.

\begin{figure}
\includegraphics{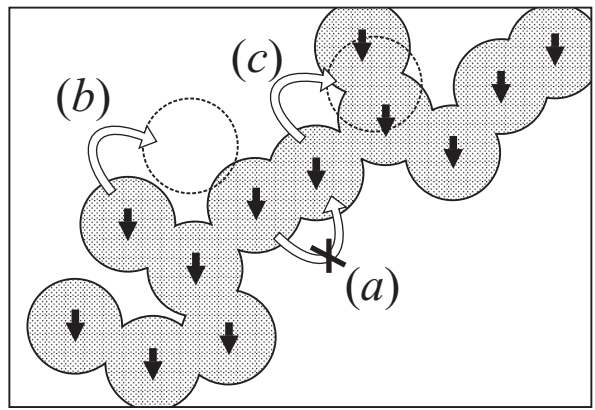}
\caption{\label{fig:polaronhop} 
  Charge transport in a DMS system with strongly localized carriers.
  The spins of localized holes are shown with black arrows, the
  regions with magnetic impurities polarized by these spins are
  dotted. Beyond the percolation threshold, the polarized regions
  coalesce, forming an infinite cluster spanning the whole sample.
  The charge transport is facilitated by holes hopping between the
  localization sites. ($a$) On-site repulsion does not allow a hole to
  hop onto a site already taken by another hole. ($b$) A hole can hop
  only to a free site, which may be outside the region polarized by
  other holes. ($c$) Hops to already polarized regions are rare at
  $T\sim T_c$.}
\end{figure}

At temperatures of the order of the ferromagnetic transition
temperature, the hole transport in the system occurs by means of
nearest-neighbor hopping. The resistivity $\rho$ of such a system
depends on temperature exponentially, 
\begin{equation}
\label{rhoT}
\rho \propto
\exp(E_{\textrm{hop}}/T)\;.
\end{equation}
The characteristic hopping activation
energy $E_{\textrm{hop}}$ has two contributions. First, there is random
level mismatch between two localization sites, produced by the
disorder. The second contribution to the hopping activation energy
comes from the interaction of a localized hole with neighboring
magnetic impurities.  This interaction leads to polarization of
impurity spins by the hole spin and lowers the energy of the resulting
bound magnetic polaron. The energy associated with this polarization
(polaron's ``binding energy'') is given by
\begin{equation}
E_{\textrm{pol}} = 8 \pi J_0 a_0^3 n_{\textrm{i}}\;,
\label{polaronEnergy}
\end{equation}
provided $r_{\textrm{pol}} > a_0$, which is satisfied at temperatures
of the order and below the ferromagnetic transition temperature $T_c$.
When a hole hops to another localization site, it abandons the region
it has polarized and lands among non-polarized impurities unless this
new localization site is within another bound magnetic polaron. The
``binding energy'' $E_{\textrm{pol}}$ of a polaron is much less than
the characteristic energy level mismatch $E_{\textrm{hop}}^{(0)} \sim
t_0$ at different localization sites. Therefore this ``binding
energy'' will not have a noticeable effect on the hopping trajectory
of a hole.  At temperatures close to the ferromagnetic transition
temperature $T_c$, while the bound magnetic polarons barely touch each
other, it means that most of the hops will be to unpolarized regions,
see Fig.~\ref{fig:polaronhop}. Therefore, the characteristic hopping
activation energy will acquire an addition term $E_{\textrm{pol}}$,
\begin{equation}
\label{highThop}
E_{\textrm{hop}}^{\vphantom{(0)}} = E_{\textrm{hop}}^{(0)} + E_{\textrm{pol}}
\ \textrm{at}\ T\lesssim T_c\;.
\end{equation}

At low enough temperatures, however, the infinite cluster of bound
magnetic polarons will cover the whole sample, so wherever a hole
hops, it will land in a region which is already polarized in a optimal
way, so the impurity polarization will not have any contribution to
the hopping activation energy. Our Monte-Carlo studies show that the
characteristic temperature $T_{\textrm{cover}}$ at which the infinite
cluster covers most of the sample corresponds to
$r_{\textrm{pol}}(T_{\textrm{cover}}) \sim 2 r_{\textrm{pol}}(T_c)$,
which yields
\begin{displaymath}
T_{\textrm{cover}}\sim T_c
\exp\left(-\frac{0.86}{\left(a_0^3n_{\textrm{h}}\right)^\frac{1}{3}}\right)
\ll T_c\;. 
\end{displaymath}
Thus
\begin{equation}
\label{lowThop}
E_{\textrm{hop}}^{\vphantom{(0)}} = E_{\textrm{hop}}^{(0)}
\ \textrm{at}\ T\lesssim T_{\textrm{cover}}\ll T_c\;.
\end{equation}

\begin{figure}
\includegraphics{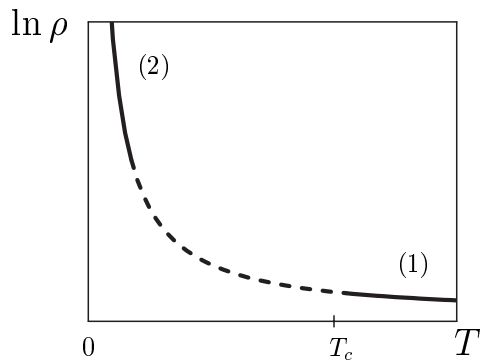}
\caption{\label{fig:polaronres} 
  Sketch of the temperature dependence of resistivity in a DMS system
  with strongly localized carriers. The hopping activation energy
  $E_{\textrm{hop}}$ is given by Eqs.~(\protect\ref{highThop}) and
  (\protect\ref{lowThop}) for regions (1) and (2) respectively, with
  the transition between the regions being sample-dependent. Even
  though $E_{\textrm{hop}}$ is lower is region (2), the temperature
  decrease needed to reduce it is large enough for $\rho(T)$ to remain
  monotonic.} 
\end{figure}

The concrete scenario of the infinite cluster spreading over the
transport paths highly depends on the sample, because even though
these two random networks are not completely independent, they
generally do not coincide. Since the size of a bound magnetic polaron
grows logarithmically slowly as the temperature goes down,
sample-dependent variations in the mutual arrangement of the transport
paths and the infinite cluster lead to strong variations in the
temperatures at which the transition from the high-temperature
[Eq.~(\ref{highThop})] to the low-temperature [Eq.~(\ref{lowThop})]
exponent occurs and in the precise functional form of $\rho(T)$ during
this transition. However, since the transition is slow, the overall
temperature dependence $\rho(T)$ of the resistivity must be monotonic,
because even though magnetization reduces the hopping barrier, the
decrease in temperature make it harder for electrons to get activated
to overcome it.

External magnetic field $B$ applied to the sample polarizes the impurities
all over the sample, with the average value of the impurity
polarization given by
\begin{equation}
\label{siofB}
\left\langle S_z \right\rangle = 
S\mathcal{B}_S
\left(\frac{g_{\textrm{i}}\mu_B B}{k_B T}\right)
\end{equation}
where $\mu_B$ is the Bohr magneton,
\begin{equation}
\label{brilloin}
\mathcal{B}_s(x) \equiv
\frac{2s+1}{2s}\coth\frac{2s+1}{2s}x-\frac{1}{2s}\coth\frac{1}{2s}x
\end{equation}
is the Brillouin function, and the direction of the $z$ axis is chosen
to be parallel to the magnetic field. As the result of this
polarization, the impurities around the localization site a hole hops
into are not entirely uncorrelated with those around the site it hops
from. The ``polaron binding energy'' (\ref{polaronEnergy}) is
therefore reduced by
\begin{equation}
\Delta E_{\textrm{pol}}(B) = - 8 \pi J_0 a_0^3 n_{\textrm{i}} \mathcal{B}_S
\left(\frac{g_{\textrm{i}}\mu_B B}{k_B T}\right)\;,
\label{deltapolaronEnergy}
\end{equation}
and the resulting magnetic-field dependence of the resistivity is
given by 
\begin{equation}
\label{rhoB}
\rho(B)=\rho\vert_{B=0} \exp\left[
- \frac{8 \pi J_0 a_0^3 n_{\textrm{i}}}{k_BT} \mathcal{B}_S
\left(\frac{g_{\textrm{i}}\mu_B B}{k_B T}\right)
\right]\;,
\end{equation}
as it follows from Eqs.~(\ref{rhoT}), (\ref{highThop}), and
(\ref{deltapolaronEnergy}). One can see that the resistivity decreases
as the applied magnetic filed grows, in agreement with the
experimental results.\cite{Matsukura98}

\section{Charge carriers localized within clusters}
\label{sec:clusters}

The previous section addressed transport in magnetic semiconductors
with strongly localized charge carriers. ``Strongly localized'' in
that case meant that a charge carrier is pinned to some point (a pair
of close impurities or some other defect in crystal structure), with
the carrier localization radius being smaller than the characteristic
distance between pinning centers. Coulomb repulsion prevents a carrier
from entering a localization site already taken by another carrier.
This regime takes place when concentrations $n_{\textrm{i}}$ of
magnetic impurities and $n_{\textrm{h}}$ of charge carriers are small.
For such small $n_{\textrm{i}}$ and $n_{\textrm{h}}$, ferromagnetic
transition temperature $T_c$ is typically of the order of few Kelvins
for realistic parameters.  For practical applications, however, it is
desirable to have $T_c$ higher than few Kelvins, which implies higher
$n_{\textrm{i}}$ and $n_{\textrm{h}}$, so one may naturally wonder
what happens to the charge transport when the magnetic impurity
concentration is so high that the assumptions of
Sec.~\ref{sec:strongly} are no longer applicable.

If the positions of impurities are uncorrelated, charge carriers
become delocalized when parameter $n_{\textrm{i}} a_0^3$ exceeds some
critical value of the order of unity. The analytical description of
this transition is very challenging, and hardly any quantitative
results can be obtained. We postpone the discussion of this case until
the next section. In this section we consider the case when magnetic
impurities are grouped into clusters. This arrangement is not
improbable at all, taking into account significant differences between
the non-magnetic atoms of the host lattice and the magnetic dopants.
In fact there are theoretical indications\cite{VonOppen2002} that the
impurities \emph{should} group into clusters, due to electrostatic
interactions during the growth process.

For the parameter region in the vicinity of $n_{\textrm{i}} a^3_B\sim
1$ clustering of impurities means that a hole will be able to move
freely within a cluster, but not between clusters, i.e. the
characteristic size of a hole wave function will be of the order of a
cluster size rather than of the size of a wave function of a hole
localized at an isolated impurity or an impurity pair. Because of the
hole wave function being spread over many impurities, the Coulomb
energy which repels a hole from a site occupied by another hole is
reduced as compared to the case of strongly localized holes considered
in Sec.~\ref{sec:strongly}. If the cluster size is large enough,
several holes may enter one cluster, and the characteristic magnitude
of the Coulomb energy in a cluster at the Fermi level is
\begin{equation}
\label{clusterCoul}
E_{\textrm{Coul}} \sim U \frac{N_{\textrm{h}}}{N_{\textrm{i}}}\;,
\end{equation}
where $N_{\textrm{h}}$ and $N_{\textrm{i}}$ are the number of
impurities in the cluster and the number of holes sitting in it,
respectively. Since $n_{\textrm{i}} \gg n_{\textrm{h}}$, for a typical
cluster we must have $N_{\textrm{i}} \gg N_{\textrm{h}}$, and,
therefore $E_{\textrm{Coul}} \ll U$.

The activation energy $\tilde{E}_{\textrm{hop}}$ for hole
hopping between the clusters contains contributions from the Coulomb-
and disorder-induced level mismatch between the clusters and from the
exchange interaction between the holes hopping between the clusters
and the impurities forming these clusters.  The first contribution
stems mainly from the Coulomb repulsion between the holes in clusters.
The characteristic value of the gap between the highest occupied and
the lowest unoccupied levels in a cluster is given by
Eq.~(\ref{clusterCoul}). The relative positions of these levels in
different clusters are completely random, so a characteristic value of
the mismatch between the highest occupied and lowest unoccupied levels
in two neighboring clusters will also be of the order of
$E_{\textrm{Coul}}$ given by Eq.~(\ref{clusterCoul}). The
random mismatch $E_{\textrm{sp}}$ between two single-particle levels
in two clusters also plays some role, but the characteristic value of
this mismatch
\begin{equation}
\label{Emism}
E_{\textrm{sp}} \sim \frac{t_0}{N^{2/3}_{\textrm{i}}N^{1/3}_{\textrm{c}}}
\end{equation}
should be much less than $E_{\textrm{Coul}}$ for realistic systems.

The second contribution to the hopping activation energy comes, as it
was stated above, from the exchange interaction between holes and
impurities. At high temperatures, spins of impurities and holes in a
cluster are not polarized, and there is no contribution to the hopping
activation energy due to exchange. At lower temperatures, however,
exchange interaction between magnetic impurities and holes leads to
ferromagnetic transition. The mechanics of the latter is as follows:
first, exchange interaction between holes and impurities in clusters
they occupy lead to ferromagnetic transition within clusters.  The
exchange interaction between the clusters is exponentially small,
$\propto \exp (-r_{\textrm{cl}}/a_0)$, where $r_{\textrm{cl}}$ is the
characteristic distance between the clusters. Therefore, as the
temperature is lowered and ferromagnetic transition occurs within
clusters, the magnetic moments of different clusters are still
uncorrelated. At even lower temperatures, the correlation between the
clusters is established, and the macroscopic ferromagnetic transition
occurs. At high temperatures when the even clusters are not polarized
a hole hopping from a cluster to another one experiences no change of
the exchange energy. However at temperatures close to the ferromagnetic
transition temperature the clusters are inevitably polarized, and
this creates an additional energy barrier for a hole hopping between
two uncorrelated clusters.

While the ferromagnetic transition in a cluster does not imply full
polarization of all impurities in it, at the temperatures of the order
of the ferromagnetic transition temperature all the impurities must be
polarized, and the energy associated with the exchange interaction of
holes in a cluster with all the impurities in it equals
\begin{equation}
\label{lfejg}
\tilde{E}_{\textrm{exch}} = J_0\;.
\end{equation}
When two clusters are not correlated, a hole hopping from one cluster to
the other one experiences the change of potential of the order of 
$E_{\textrm{exch}}$, so the expression for the characteristic hopping
activation energy reads
\begin{equation}
\label{EhopClust}
\tilde{E}_{\textrm{hop}} = \tilde{E}_{\textrm{Coul}} +
\tilde{E}_{\textrm{exch}}\ \textrm{at}\ T_c\lesssim T\;.
\end{equation}

\begin{figure}
\includegraphics{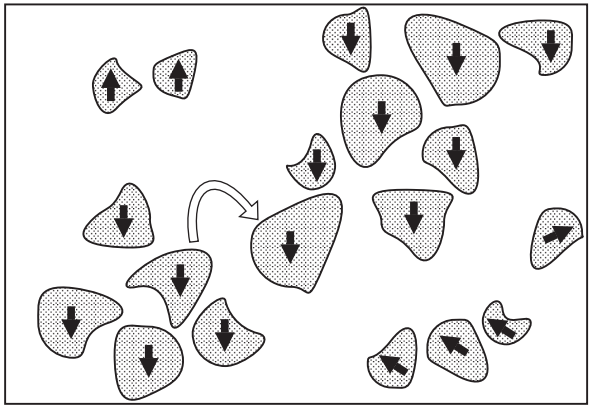}
\caption{\label{fig:clusterhop} 
  Charge transport in a DMS system with clustered magnetic impurities.
  The picture refers to the temperatures low enough for the
  ferromagnetic transition to occur within clusters, whose spins are
  shown with black arrows. Holes hop between the clusters preferring
  shorter gaps, so the percolation paths coincide with transport paths.}
\end{figure}

\begin{figure}
\includegraphics{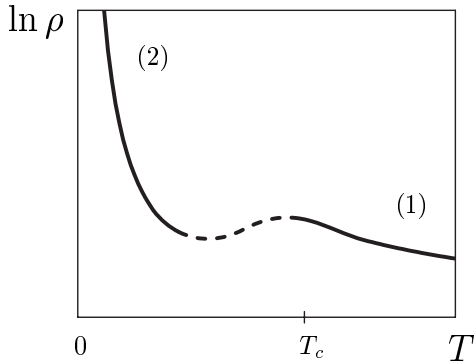}
\caption{\label{fig:clusterres} 
  Sketch of the temperature dependence of resistivity in a DMS system
  with clustered impurities. Since percolation and transport paths
  coincide, ferromagnetic transition leads to immediate decrease in
  the hopping energy, which may result in non-monotonic temperature
  behavior of the sample resistivity. }
\end{figure}

At temperatures below $T_c$, the correlation between the cluster spins
is established, and a hole hopping into another cluster experiences
exactly the same exchange potential as in the cluster it has left.
This reduction of the exchange contribution to the hopping activation
energy from $J_0$ to zero (roughly speaking) occurs while temperature
$T$ is still of the order of $T_c$. Therefore, the exponent in
Eq.~(\ref{rhoT}) may actually become smaller as the temperature goes
down past $T_c$, provided the second term in Eq.~(\ref{EhopClust}),
which vanishes when the spins of clusters become aligned, is larger
than the first, temperature-independent term, that is 
\begin{equation}
\label{ohoho}
U \frac{N_{\textrm{h}}}{N_{\textrm{i}}} < J_0
\end{equation}

The external magnetic field applied to a sample lowers its resistance,
similarly to the case discussed in the previous section. However, now
the reduction of the resistivity in the systems is due to polarization
of impurity clusters as a whole, not of individual impurities, since
all impurities in a cluster have their spins aligned in the same
direction. Since polarization of a cluster by magnetic field of a
given magnitude strongly depends on the size of the cluster, and since
the sizes of impurity clusters are random with possibly wide
distribution, the quantitative dependence of resistivity on the
applied magnetic field is highly sample-dependent.

\section{Charge carriers on the verge of delocalization}
\label{sec:verge}

At sufficiently high concentrations of impurities and holes, some of
the holes are delocalized even in the limit $T\to 0$. The
ferromagnetic transition in systems of this class does not have
percolation nature, unless the fraction of delocalized carriers is
very small. With physics of their ferromagnetic transition and charge
transport being entirely different from that of the systems with
localized charge carriers, these systems are beyond the scope of the
current paper. In this section, we consider systems which are close to 
localization threshold, but still not beyond it. In the ground state of 
such a system, all charge carriers are localized within some finite
regions, and the resistivity of the sample goes to infinity as $T\to
0$. However, because of the small separation between the Fermi level
and the mobility edge, finite temperature excites some holes into
delocalized states. It is hardly possible to make any quantitative
statements about magnetization and resistivity of the system, but
still some qualitative statements can be made with the help of
intuition we have developed dealing with the case of the previous
section.

\begin{figure}
\includegraphics{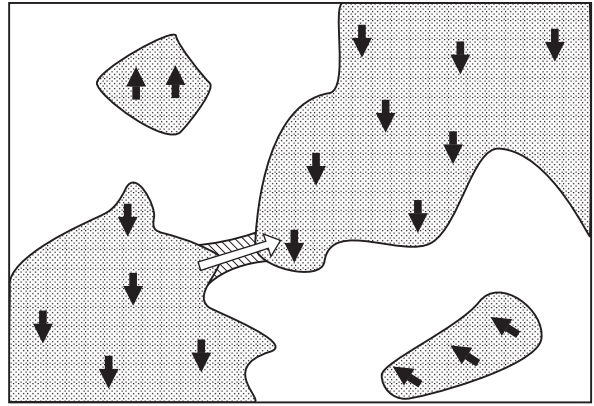}
\caption{\label{fig:verge} 
  Charge transport in a DMS system with charge carriers on the verge of
  being delocalized. Dotted areas show ``puddles'' of charge carriers,
  with impurity spins correlated across a puddle even at temperatures
  above the ferromagnetic transition. Gaps between puddles are bridged
  by thermally excited delocalized states (hatched region).}
\end{figure}

Since, despite some holes being thermally excited into delocalized
states, many holes are still localized, the effective exchange
interaction between the magnetic impurities, which is induced by
holes, is still highly inhomogeneous across the sample. There will be
``puddles'' filled with holes, where interimpurity exchange
concentration is strong, and relatively hole-free regions, which are
covered only by tails of hole wave functions decaying away from these
puddles, see Fig.~\ref{fig:verge}. Qualitatively, this picture is
similar to that of Sec.~\ref{sec:clusters} shown in
Fig.~\ref{fig:clusterhop}, with the difference that in the case
considered in the present section transport paths go mostly along
hole-filled regions with fewer gaps on the way. These gaps are bridged
by the delocalized states, so there is no need for hopping.

Similarly to the case of previous sections, it is regions with higher
hole density that undergo (local) ferromagnetic transition first, with
ferromagnetic correlations across the sample established at lower
temperatures. When two neighboring ``hole puddles'' have the different
spin orientation, a delocalized hole going from one puddle to the
other encounters an additional barrier, which can partially reflect it
back or even entirely prevent it from entering the puddle. When the
spins of two puddles oriented in the same direction, as it is the case
at the temperatures below the ferromagnetic transition temperature, no
such problem arise, and a delocalized hole move freely from a puddle
to a puddle across the sample.

We thus have two competing phenomena that occur as the temperature
goes down. On one hand, at lower temperatures we have fewer holes
excited to delocalized states. On the other hand ferromagnetic
transition occuring as temperature goes down makes it easier for the
hole to travel across the sample. The interplay of these two phenomena
determines the temperature dependence of resistivity. Due to the
extremely complex nature of hole wave functions close to the mobility
edge, we are unable to provide any qualitative results. However, with
the system under consideration being in some sense an extreme case of
the system consider in the previous section, we may deduce that the
non-monotonic behavior of resistivity must be possible, and that the
non-monotonicity becomes more pronounced as the Fermi level of the
holes approaches the mobility edge, i.e. when the concentrations of
holes and impurities become higher. Similarly to the previous cases,
magnetic field applied to the system would align the spins of hole
puddles and enhance the transport.

\section{Conclusion}

We have considered correlations between ferromagnetic percolation
transition and charge transport in magnetic semiconductors with
localized charge carriers. We found that ferromagnetic transition
alone would enhance charge transport, but the temperature decrease,
which is needed for the transition, may completely eliminate these
gains. In a system with strongly localized charge carriers
(Sec.~\ref{sec:strongly}), decrease in the hopping rate of the charge
carriers due to temperature decrease overcomes decrease in the hopping
activation energy due to ferromagnetic transition, and the temperature
dependence of the resistivity is always monotonic. At higher
concentrations of magnetic impurities and charge carriers, the
situation may be the opposite, and resistivity may experience a dip as
the temperature goes down past the ferromagnetic transition
temperature (Secs.~\ref{sec:clusters}, \ref{sec:verge}). The higher
the concentrations of the impurities and carriers are, the more
pronounced the dip is. In all cases considered in the paper, the
resistivity of the system decreases if an external magnetic field is
applied.

It is important to point out that our theoretical DMS transport
results are in qualitative agreement with the available experimental
transport results in the localized insulating regime, where the
resistivity increases strongly (exponentially) with temperature as
$T\to0$. In particular, deep into the insulating regime, far away from
the metal-insulator transition point, the experimentally
measured\cite{Matsukura98,Tanaka2003} resistivity rises
monotonically with decreasing temperature similar to our results
(Fig.~\ref{fig:polaronres}) presented in Sec.~\ref{sec:strongly} for
strongly localized charge carrier system.  On the other hand, close to
the metal-insulator transition (but still on the insulating side)
transport experiments indeed
observe\cite{Matsukura98,Tanaka2003} a striking
temperature-dependent nonmonotonicity in the measured resistivity as
we find in our theory for ``not-so-strongly'' localized charge
carriers (Sec.~\ref{sec:clusters}, Fig.~\ref{fig:clusterres}).
Unfortunately, the insulating DMS ferromagnetic regime has not yet
been extensively studied experimentally (compared with the metallic
DMS regime with optimal $T_c$ values) although our view is that this
regime is as important and as interesting as the metallic regime in
terms of the development of our understanding of DMS physics. We urge
that more experimental work done in the insulating ferromagnetic
regime (and on the metal-insulator transition itself) for us to
develop a better and more quantitative understanding of transport and
magnetization phenomena in DMS systems.

This work was supported by DARPA and US-ONR.

\bibliography{transper}

\begin{thebibliography}{15}
\expandafter\ifx\csname natexlab\endcsname\relax\def\natexlab#1{#1}\fi
\expandafter\ifx\csname bibnamefont\endcsname\relax
  \def\bibnamefont#1{#1}\fi
\expandafter\ifx\csname bibfnamefont\endcsname\relax
  \def\bibfnamefont#1{#1}\fi
\expandafter\ifx\csname citenamefont\endcsname\relax
  \def\citenamefont#1{#1}\fi
\expandafter\ifx\csname url\endcsname\relax
  \def\url#1{\texttt{#1}}\fi
\expandafter\ifx\csname urlprefix\endcsname\relax\def\urlprefix{URL }\fi
\providecommand{\bibinfo}[2]{#2}
\providecommand{\eprint}[2][]{\url{#2}}

\bibitem[{\citenamefont{Ohno et~al.}(1996)\citenamefont{Ohno, Shen, Matsukura,
  Oiwa, Endo, Katsumoto, and Iye}}]{Ohno96}
\bibinfo{author}{\bibfnamefont{H.}~\bibnamefont{Ohno}},
  \bibinfo{author}{\bibfnamefont{A.}~\bibnamefont{Shen}},
  \bibinfo{author}{\bibfnamefont{F.}~\bibnamefont{Matsukura}},
  \bibinfo{author}{\bibfnamefont{A.}~\bibnamefont{Oiwa}},
  \bibinfo{author}{\bibfnamefont{A.}~\bibnamefont{Endo}},
  \bibinfo{author}{\bibfnamefont{S.}~\bibnamefont{Katsumoto}},
  \bibnamefont{and} \bibinfo{author}{\bibfnamefont{Y.}~\bibnamefont{Iye}},
  \bibinfo{journal}{Appl. Phys. Lett.} \textbf{\bibinfo{volume}{69}},
  \bibinfo{pages}{363} (\bibinfo{year}{1996}).

\bibitem[{\citenamefont{Kaminski and {Das Sarma}}(2002)}]{KaminskiDasSarma2002}
\bibinfo{author}{\bibfnamefont{A.}~\bibnamefont{Kaminski}} \bibnamefont{and}
  \bibinfo{author}{\bibfnamefont{S.}~\bibnamefont{{Das Sarma}}},
  \bibinfo{journal}{\prl} \textbf{\bibinfo{volume}{88}},
  \bibinfo{pages}{247202} (\bibinfo{year}{2002}).

\bibitem[{\citenamefont{Matsukura et~al.}(1998)\citenamefont{Matsukura, Ohno,
  Shen, and Sugawara}}]{Matsukura98}
\bibinfo{author}{\bibfnamefont{F.}~\bibnamefont{Matsukura}},
  \bibinfo{author}{\bibfnamefont{H.}~\bibnamefont{Ohno}},
  \bibinfo{author}{\bibfnamefont{A.}~\bibnamefont{Shen}}, \bibnamefont{and}
  \bibinfo{author}{\bibfnamefont{Y.}~\bibnamefont{Sugawara}},
  \bibinfo{journal}{\prb} \textbf{\bibinfo{volume}{57}} (\bibinfo{year}{1998}).

\bibitem[{\citenamefont{Potashnik et~al.}(2002)\citenamefont{Potashnik, Ku,
  Mahendiran, Chun, Wang, Samarth, and Schiffer}}]{Potashnik02}
\bibinfo{author}{\bibfnamefont{S.~J.} \bibnamefont{Potashnik}},
  \bibinfo{author}{\bibfnamefont{K.~C.} \bibnamefont{Ku}},
  \bibinfo{author}{\bibfnamefont{R.}~\bibnamefont{Mahendiran}},
  \bibinfo{author}{\bibfnamefont{S.~H.} \bibnamefont{Chun}},
  \bibinfo{author}{\bibfnamefont{R.~F.} \bibnamefont{Wang}},
  \bibinfo{author}{\bibfnamefont{N.}~\bibnamefont{Samarth}}, \bibnamefont{and}
  \bibinfo{author}{\bibfnamefont{P.}~\bibnamefont{Schiffer}},
  \bibinfo{journal}{\prb} \textbf{\bibinfo{volume}{66}},
  \bibinfo{pages}{012408} (\bibinfo{year}{2002}).

\bibitem[{\citenamefont{K{\"o}nig et~al.}()\citenamefont{K{\"o}nig, Schliemann,
  Jungwirth, and MacDonald}}]{KonigReview01}
\bibinfo{author}{\bibfnamefont{J.}~\bibnamefont{K{\"o}nig}},
  \bibinfo{author}{\bibfnamefont{J.}~\bibnamefont{Schliemann}},
  \bibinfo{author}{\bibfnamefont{T.}~\bibnamefont{Jungwirth}},
  \bibnamefont{and} \bibinfo{author}{\bibfnamefont{A.~H.}
  \bibnamefont{MacDonald}}, \bibinfo{howpublished}{cond-mat/0111314 and
  references therein}.

\bibitem[{gri()}]{griffiths}
\bibinfo{howpublished}{This scenario is relevant to the Griffiths phase physics
  [R. B. Griffiths, \prl \textbf{23}, 17 (1969)], whose application to diluted
  magnetic semiconductors is considered in our paper V. M. Galitski, A.
  Kaminski, and S. Das Sarma, cond-mat/0306488.}

\bibitem[{\citenamefont{Mayr et~al.}(2002)\citenamefont{Mayr, Alvarez, and
  Dagotto}}]{Dagotto02}
\bibinfo{author}{\bibfnamefont{M.}~\bibnamefont{Mayr}},
  \bibinfo{author}{\bibfnamefont{G.}~\bibnamefont{Alvarez}}, \bibnamefont{and}
  \bibinfo{author}{\bibfnamefont{E.}~\bibnamefont{Dagotto}},
  \bibinfo{journal}{\prb} \textbf{\bibinfo{volume}{65}},
  \bibinfo{pages}{241202} (\bibinfo{year}{2002}).

\bibitem[{\citenamefont{Yang and MacDonald}()}]{YangMacDonald02}
\bibinfo{author}{\bibfnamefont{S.-R.~E.} \bibnamefont{Yang}} \bibnamefont{and}
  \bibinfo{author}{\bibfnamefont{A.~H.} \bibnamefont{MacDonald}},
  \eprint{cond-mat/0202021}.

\bibitem[{\citenamefont{Korenblit et~al.}()\citenamefont{Korenblit, Shender,
  and Shklovskii}}]{KorenblitEtAl73}
\bibinfo{author}{\bibfnamefont{I.~Y.} \bibnamefont{Korenblit}},
  \bibinfo{author}{\bibfnamefont{E.~F.} \bibnamefont{Shender}},
  \bibnamefont{and} \bibinfo{author}{\bibfnamefont{B.~I.}
  \bibnamefont{Shklovskii}}, \bibinfo{howpublished}{Phys. Lett. \textbf{46A},
  275 (1973)}.

\bibitem[{\citenamefont{Wan and Bhatt}()}]{WanBhatt00}
\bibinfo{author}{\bibfnamefont{X.}~\bibnamefont{Wan}} \bibnamefont{and}
  \bibinfo{author}{\bibfnamefont{R.~N.} \bibnamefont{Bhatt}},
  \eprint{cond-mat/0009161.}

\bibitem[{\citenamefont{Nazmul et~al.}(2003)\citenamefont{Nazmul, Sugahara, and
  Tanaka}}]{Tanaka2003}
\bibinfo{author}{\bibfnamefont{A.~M.} \bibnamefont{Nazmul}},
  \bibinfo{author}{\bibfnamefont{S.}~\bibnamefont{Sugahara}}, \bibnamefont{and}
  \bibinfo{author}{\bibfnamefont{M.}~\bibnamefont{Tanaka}},
  \bibinfo{journal}{\prb} \textbf{\bibinfo{volume}{67}},
  \bibinfo{pages}{241308} (\bibinfo{year}{2003}).

\bibitem[{\citenamefont{{Das Sarma} et~al.}(2003)\citenamefont{{Das Sarma},
  Hwang, and Kaminski}}]{DasSarmaHwangKaminski2002}
\bibinfo{author}{\bibfnamefont{S.}~\bibnamefont{{Das Sarma}}},
  \bibinfo{author}{\bibfnamefont{E.}~\bibnamefont{Hwang}}, \bibnamefont{and}
  \bibinfo{author}{\bibfnamefont{A.}~\bibnamefont{Kaminski}},
  \bibinfo{journal}{\prb} \textbf{\bibinfo{volume}{67}},
  \bibinfo{pages}{155201} (\bibinfo{year}{2003}).

\bibitem[{\citenamefont{Schrieffer and Wolff}(1966)}]{SchriefferWolff66}
\bibinfo{author}{\bibfnamefont{J.~R.} \bibnamefont{Schrieffer}}
  \bibnamefont{and} \bibinfo{author}{\bibfnamefont{P.~A.} \bibnamefont{Wolff}},
  \bibinfo{journal}{Phys. Rev.} \textbf{\bibinfo{volume}{149}},
  \bibinfo{pages}{491} (\bibinfo{year}{1966}).

\bibitem[{\citenamefont{Shklovskii and Efros}(2001)}]{EfrosShklovskiiBook}
\bibinfo{author}{\bibfnamefont{B.~I.} \bibnamefont{Shklovskii}}
  \bibnamefont{and} \bibinfo{author}{\bibfnamefont{A.~L.} \bibnamefont{Efros}},
  \emph{\bibinfo{title}{Electronic properties of Doped Semiconductors}}
  (\bibinfo{publisher}{Springer-Verlag}, \bibinfo{address}{Berlin},
  \bibinfo{year}{2001}).

\bibitem[{\citenamefont{Timm et~al.}(2002)\citenamefont{Timm, Sch{\"{a}}fer,
  and {von Oppen}}}]{VonOppen2002}
\bibinfo{author}{\bibfnamefont{C.}~\bibnamefont{Timm}},
  \bibinfo{author}{\bibfnamefont{F.}~\bibnamefont{Sch{\"{a}}fer}},
  \bibnamefont{and} \bibinfo{author}{\bibfnamefont{F.}~\bibnamefont{{von
  Oppen}}}, \bibinfo{journal}{Phys. Rev. Lett.} \textbf{\bibinfo{volume}{89}},
  \bibinfo{pages}{137201} (\bibinfo{year}{2002}).

\end{thebibliography}

\end{document}